\documentclass[a4paper,12pt]{article}
\usepackage{latexsym,amsmath,amsfonts,amssymb,mathrsfs,caption,array}
\usepackage[latin1]{inputenc}
\usepackage{moreverb,graphicx,color,caption,macros,multirow,footmisc,sint}
\usepackage[numbers,sort]{natbib}
\usepackage[pdftex]{hyperref}
\hypersetup{colorlinks,linkcolor=black,filecolor=black,urlcolor=black,citecolor=black,plainpages=false,
hypertexnames=false}

\newcolumntype{C}[1]{>{\centering}m{#1}}
\begin{document}
\begin{flushright} HU-EP-10/26\\ SFB/CPP-10-38\\ DESY 10-075\end{flushright}
\begin{center}
  {\Large\bf  The running coupling of QCD with four flavors 
  }
\end{center}
\begin{figure}[!h]
$$\includegraphics[height=1cm]{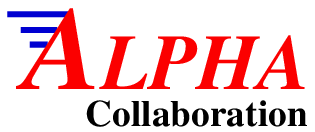}$$
\end{figure}
\begin{center}
Fatih Tekin\footnote{fatih@physik.hu-berlin.de\label{Fatih}}, Rainer Sommer\footnote{rainer.sommer@desy.de\label{Rainer}}, Ulli Wolff\footnote{uwolff@physik.hu-berlin.de\label{Ulli}}
\vskip 0.5 cm
$^{\rm \ref{Fatih},\ref{Ulli}}$ 
Institut f\"ur Physik, Humboldt Universit\"at\\
Newtonstr. 15, 12489 Berlin,\\ 
Germany
\vskip 1.5ex
$^{\rm \ref{Rainer}}$
NIC, DESY\\
Platanenallee 6, 15738 Zeuthen,\\ 
Germany
\vskip 0.5 cm
{\bf Abstract}
\vskip 0.7ex
\end{center}
We have calculated the step scaling function and the running coupling of QCD in the \SF scheme with four flavors of $\mathcal{O}(a)$ improved Wilson quarks. Comparisons of our non-perturbative results with 2-loop and 3-loop perturbation theory as well as with non-perturbative data
for only two flavors are made.
\thispagestyle{empty}
\newpage
\section{Introduction}
One ingredient of the contemporary Standard Model is QCD which is believed to describe the strong interaction between quarks and gluons. The theory has six free mass parameters for the six quark species and one coupling parameter. After fixing these free parameters of the QCD Lagrangian at some reference energy scale, the theory in principle predicts presently known effects in which only the strong interaction is involved. Since QCD possesses the property of asymptotic freedom, perturbation theory is applicable in the high energy regime ($E\gg 1\GeV$), but for low energies lattice QCD, the only known systematic non-perturbative approach, is required. Here a certain number of physical quantities, for example meson masses, have to be identified
with experimental inputs to determine the free parameters and in particular to gauge dimensionful quantities on the lattice in powers of $\GeV$. 
Only then predictions become possible. As an example, the light hadron mass
spectrum was studied by different collaborations in quenched as well as in
full QCD with up to three dynamical fermion 
species \cite{Butler:1994,spectrum:cppacs_nf0,spectrum:cppacs_nf2,Aoki:2008sm,spectrum:etmc_nf2_baryons,Durr:2008zz}. The theoretical predictions for the spectrum in full QCD are compatible with the experiments at the error-level achieved so far.
 
At a first glance the perturbative and non-perturbative formulations seem to
be disjoint and have their own domains of applicability. In a long-term
project, the ALPHA collaboration has been developing methods and tools to
connect these regimes by computing the parameters of the perturbative domain
starting from the non-perturbative formulation \cite{Luscher:1998pe}. In this
context, the step scaling function which we will discuss later plays a key
role. For different numbers of flavors, 
the step scaling function \cite{Luscher:1993gh,mbar:pap1,Takeda:2004,DellaMorte:2004bc,alpha:cppacs_nf3,Appelquist:2009ty} which can be understood as an integrated form of the QCD 
$\beta$-function and the running coupling have been examined and determined successfully (see \cite{Mason:2005zx,Gockeler:2005rv,Shintani:2008ga} for recent publications).  In this paper, we want to calculate the step scaling function and the running coupling with four flavors of $\mathcal{O}(a)$ improved Wilson quarks. For the present work, the improvement coefficient $\csw$ which is essential for the $\mathcal{O}(a)$ improvement with Wilson quarks was determined in a preceding paper \cite{Tekin:2009}. We exploit this result in the simulations that we report in the following.

The paper is organized as follows. In sections 2--5 we summarize some background about the nonperturbative definition
of a coupling constant in the \SF finite volume renormalization scheme and about the step scaling technique.
In section 6 we present our new raw data and their analysis.
We will arrive, for four flavors, at a value for the dimensionless combination $\Lambda\Lmax$.
While $\Lambda$ parameterizes the coupling at high energy, $\Lmax$ is an unambiguously defined length in the hadronic regime.
Its value in GeV remains to be determined by large volume simulations.

\section{The finite size strategy}
The fundamental parameters of QCD, i.e. the coupling and masses of the quarks, depend on a scale $\mu$.
To make contact with perturbative QCD we need to know their values for large $\mu$ in the domain of asymptotic freedom.
If such a computation is attempted on a single lattice where at the same time hadronic scales are measurable with small finite volume and discretization effects
one has to satisfy to a reasonable precision the string of inequalities
\begin{equation}\label{demandonsimulations}
 L\gg\frac{1}{m_{\pi}}\approx\frac{1}{0.14\,\GeV}\gg\frac{1}{\mu}\approx\frac{1}{10\,\GeV}\gg a.
\end{equation}
In total this implies $L/a\gg 70$ which is impossible to satisfy for some time. A way to circumvent these difficulties was 
hence proposed by L\"uscher et al. in \cite{Luscher:1991wu}. The idea is to exploit the universality of the finite volume  continuum limit
to perform intermediate renormalizations with the finite size as a renormalization scale, $\mu=1/L$.
In these steps only $L\gg a$ has to be assured to render cutoff effects small. By such simulations one can nonperturbatively
determine the change of the coupling constant in the continuum limit under scale changes from $L$ to $L/2$. By repeating such
steps we start at $\Lmax\approx\mathcal{O}(0.5\fm)$
and arrive after $k$ steps at $\mu=2^{k}/\Lmax$. With a sufficient number of steps this will be
in the perturbative regime where we can make contact with the $\Lambda$ parameter of an arbitrary scheme and with the
scheme independent renormalization
group invariant quark masses by applying perturbative formulas. If one finally succeeds in computing the precise value of
$\Lmax$ in units of a mass or decay constant, then all references to the intermediate finite volumes will have
disappeared from the final result. More details may be found in \cite{Luscher:1998pe}.

\section{Running coupling}
As mentioned before, the coupling and the quark masses are renormalization scale dependent and run with energy. Therefore the quoted values of the quark masses and the world average of $\alphas$, both in the $\MSbar$ scheme, in the Particle Physics Booklet \cite{ParticleDataGroup} are referred to a particular reference scale ($\mu\approx 2\,\GeV$ for the masses and $\mu={{M}}_{\textrm{Z}}$ for the coupling). 

From the theoretical point of view, the running of the QCD parameters is described by the renormalization group equation.
However, a physical observable $\mathscr{O}$ should have no reference to a particular renormalization scale $\mu$. This fact is expressed by the 
Callan-Symanzik equation
\begin{equation}\label{calansymanzik}
 \left[\mu\frac{\partial}{\partial\mu}+\beta(\gbar)\frac{\partial}{\partial\gbar}+\tau(\gbar)\sum_{i=1}^{\Nf}\mbar_{i}\frac{\partial}{\partial\mbar_{i}}\right]\mathscr{O}=0.
\end{equation}
In words: for any change of $\mu$ there are accompanying modifications of $\gbar$ and $\mbar_i$ such that
the physics is unchanged.
The implied scale dependence of the coupling is given by $\beta(\gbar)$ in (\ref{calansymanzik}) 
\begin{equation}\label{betafunction}
 \beta(\gbar)=\mu\frac{\partial\gbar(\mu)}{\partial\mu}.
\end{equation}
For high energies or for weak couplings, the $\beta$-function has the following asymptotic expansion
\begin{equation}\label{asymptotic}
 \beta(\gbar)\stackrel{\gbar\to 0}{=}-\gbar^{3}\left[b_{0}+b_{1}\gbar^{2}+b_{2}\gbar^{4}+\dots\right].
\end{equation}
The first two (1- and 2-loop) coe\ffi cients $b_{0}$ and $b_{1}$ in (\ref{asymptotic}) are 
scheme independent
\begin{align}
 b_{0}&=\frac{1}{(4\pi)^{2}}\left(11-\frac{2}{3}\Nf\right),\\
 b_{1}&=\frac{1}{(4\pi)^{4}}\left(102-\frac{38}{3}\Nf\right).
\end{align}
The 3-loop coefficient does depend on the scheme and
  in the \SF scheme, which we will need later, $b_{2}$ has been given in \cite{Bode:1999sm}
\begin{equation}\label{b2}
 b_{2}=\frac{1}{(4\pi)^3}\left[0.483(7)-0.275(5)\Nf+0.0361(5)\Nf^2-0.00175(1)\Nf^3\right].
\end{equation}
We restrict ourselves to mass-independent schemes where all renormalization conditions are imposed at vanishing quark masses. 
Examples are the $\MSbar$ scheme of dimensional regularization and the \SF
scheme used here. For $\Nf\leq 16$, the $\beta$-function (\ref{asymptotic}) is
negative at weak coupling 
and the integration of (\ref{betafunction}) results in a coupling which
decreases with increasing energy. 
In other words, the quarks behave like free particles in the high energy
regime (asymptotic freedom). But for $\Nf>16$, there is a sign change
and the property of asymptotic freedom is lost. 
The relation of $\gbar(\mu)$ to the $\Lambda$ parameter of QCD is given by the following solution of the Callan-Symanzik equation
 \begin{eqnarray}\nonumber
 \Lambda&=&\mu\left[b_{0}\gbar^2(\mu)\right]^{-\frac{b_{1}}{2b_{0}^2}}\exp\left\{-\frac{1}{2b_{0}\gbar^{2}(\mu)}\right\}\times\\\label{Lambda}
& &\exp\left\{-\int_{0}^{\gbar(\mu)}dx\left[\frac{1}{\beta(x)}+\frac{1}{b_{0}x^{3}}-\frac{b_{1}}{b_{0}^{2}x}\right]\right\}.
\end{eqnarray}
As is well-known, $\Lambda$ is scheme-dependent but the transformation to other schemes follows from the relation between their couplings
at one loop accuracy.
\section{Coupling in the SF scheme}
The finite-volume scheme which we use for our simulations is the \SF (SF) scheme. In the following, we will only give a brief 
reminder of the main features and properties of this scheme which are discussed in detail in many papers, for example \cite{Luscher:1992an,Sint:1993un,Sint:1995rb,Sint:1995ch,Bode:1999sm}. 

The original proposal was made in \cite{Luscher:1992an}. The construction was guided by several design criteria.
The coupling constant was to be defined independently of perturbation theory, but its evaluation had to be manageable
both in lattice perturbation theory (also beyond one loop) and by numerical simulation. Moreover, small lattice artefacts
were demanded. These requirements are not easy to fulfill and have led to the \SF which is the Euclidean propagation kernel of a field configuration $C$ at time $x_{0}=0$ to another field configuration $C'$ at time $x_{0}=T$. The free energy or effective action  
$\Gamma$ of such a system is given by
\begin{equation}\label{schroedingerpartition}
 Z[C',C]=\exp\{-\Gamma\}=\int_{\textrm{V}}\mathscr{D}[U,\Psibar,\Psi]\exp\left\{-S[U,\Psibar,\Psi]\right\}.
\end{equation}
The action $S$ is defined as in \cite{Tekin:2009}. The choice of $C_{k}$ and $C'_{k}$ of the boundary gauge fields
\begin{align}\label{ck}
 U(x,k)|_{x_{0}=0}&=\exp\{aC_{k}\},\\\label{ckprime}
 U(x,k)|_{x_{0}=T}&=\exp\{aC'_{k}\}
\end{align}
is largely arbitrarily. After some optimization with regard to lattice artefacts constant Abelian fields turned out to be appropriate. 
We will also adopt this choice of boundary fields parameterized by the scale $L$
and two dimensionless real parameters $\eta$ and $\nu$
\cite{Luscher:1993gh}. Numerical simulations of the \SF showed that the choice
$\nu=0$ leads to small statistical errors for the coupling in
simulations. Therefore, we will also set $\nu$ to zero. 

The \SF coupling is now defined by the response of $\Gamma$ to a variation of the boundary fields 
around point 'A' of \cite{Luscher:1993gh}
via the parameter $\eta$,
\begin{equation}
 \Gamma' = \left.\frac{\partial\Gamma}{\partial\eta}\right|_{\eta=0}=\frac{k}{\gbar^2},
\end{equation}
where $k$ is a normalization constant chosen such that the perturbative expansion of $\Gamma'$ 
begins with the bare coupling at tree level \cite{Luscher:1993gh}.
It should be noticed that the only external scale which appears in the definition of the coupling is the box size $L$, i.e. recursive finite size techniques can be used for the investigation of the evolution of the coupling. 

The key quantity $\Gamma'$ in the definition of the coupling is an observable which can be calculated easily through Monte Carlo simulations. Taking the derivative of (\ref{schroedingerpartition}) results in
\begin{align}
 \Gamma'&=-\frac{\partial}{\partial\eta}\ln\left\{\int\mathscr{D}[U,\Psibar,\Psi]\exp\{-S[U,\Psibar,\Psi]\}\right\}
 \\
&=\frac{1}{Z}\int\mathscr{D}[U,\Psibar,\Psi]\left(\frac{\partial\Sg}{\partial\eta}+\frac{\partial\Sf}{\partial\eta}\right)\exp\{-S[U,\Psibar,\Psi]\}
\\\label{gaugeandfermion}
 &=\left\langle\frac{\partial\Sg}{\partial\eta}\right\rangle+\left\langle\frac{\partial\Sf}{\partial\eta}\right\rangle.
\end{align}
Explicit expressions for both expectation values in (\ref{gaugeandfermion}) can be found in \cite{Sommer:2006sj,DellaMorte:2004bc}. The calculation of the renormalized coupling $\gbar^2$ involves expectation values of a local operator and no correlation functions. Therefore the numerical evaluation on a computer is straight forward once configurations are available.
The relation to the QCD coupling in the \SF scheme $\alphaSF$ is given by
\begin{equation}\label{alphasf}
 \alphaSF(\mu)=\frac{\gbar^2(L)}{4\pi},\quad \mu=1/L.
\end{equation}

To complete the definition of the coupling, also the boundary conditions for
the quark fields have to be specified. We follow exactly
\cite{Sint:1995ch}. In particular the angle entering the spatial 
periodicity of the quarks is chosen as $\theta=\pi/5$ since this value is
advantageous for the numerical simulations \cite{Sint:1995ch,DellaMorte:2004bc}.

\section{The step scaling function\label{ssf_cont}}
The concept of the \textit{step scaling function} which was introduced in \cite{Luscher:1991wu} has proven to be a very useful recursive technique to scale the coupling to high energies. As we discussed before, in our finite-volume scheme, the energy scale $\mu$ is identified with $L^{-1}$. Hence the renormalization group function $\beta$ (\ref{betafunction}) describes how the coupling changes if the box size is varied infinitesimally. The step scaling function $\sigma (s,u)$, in comparison, gives then a description how the coupling behaves when the box size $L$ is scaled by a factor $s$
\begin{equation}\label{stepscaling}
 \gbar^2(sL)=\sigma (s,\gbar^2(L)).
\end{equation}
The function $\sigma (s,u)$ can be regarded as an integrated form of
the renormalization group $\beta$-function. With the help of the step scaling
function (\ref{stepscaling}) the coupling can be traced to scales
$2^{-k}\Lmax$ (small box sizes, high energies) starting with an initial value
$L=\Lmax$. The value $s=2$ is commonly used
\cite{Luscher:1991wu,DellaMorte:2004bc,Gehrmann:2001yn} in the application of
$\sigma (s,u)$ and we will also make this choice (from now on $s=2$ and we will drop the argument $s$). The relation between the renormalization group $\beta$-function and the step scaling function $\sigma(\gbar^2(L))$ is given by
\begin{equation}\label{beta23loop}
 -2\ln (2)=\int_{u}^{\sigma(u)}\frac{dx}{\sqrt{x}\beta(\sqrt{x})}.
\end{equation}
For small values of the coupling $u$, the step scaling function has the following perturbative expansion
\begin{equation}\label{sigmaseries}
 \sigma (u)=u+s_{0}u^2+s_{1}u^3+s_{2}u^4+\dots
\end{equation}
where the coefficients are given by 
\begin{align}
 s_{0}&=2b_{0}\ln (2),\\
 s_{1}&=\left[2b_{0}\ln (2)\right]^2+2b_{1}\ln (2),\\
 s_{2}&=\left[2b_{0}\ln (2)\right]^3+10b_{0}b_{1}\left[\ln (2)\right]^2+2b_{2}\ln (2).
\end{align}
Both eq. (\ref{beta23loop}) with the truncated $\beta$ function as well as the expansion (\ref{sigmaseries}) can be used as the perturbative approximation of the step scaling function. They differ from each other by higher order terms.  
We take the first option when we compare our Monte Carlo results
with perturbation theory.

To study $\sigma(u)$ numerically one starts with choosing several lattice sizes $L/a$ and tuning the bare coupling $\gbare^2$ and the hopping parameter $\kappa$ in such a way that the value of the renormalized coupling $\gbar^2(L)$ reaches some chosen value $\gbar^2(L)=u$ and the quark mass vanishes since we are in a massless scheme. The next step is to take $L/a\to 2L/a$ and simulate at the same bare parameters. The obtained coupling $\gbar^2(2L)$ from the latter simulations is a lattice approximation $\Sigma (u,a/L)$ of the continuum step scaling function $\sigma(u)$. An extrapolation to the continuum of the data points at the same coupling $u$ but growing $L/a$  then leads to one value of the continuum function $\sigma(u)$. The procedure is repeated until a suitable range of $u$ is covered. An appropriate functional description of the continuum step scaling function can 
be given in the end in the form of a suitable fit-function that interpolates the data.

The lattice approximation $\Sigma(u,a/L)$ of the step scaling function contains remnant lattice effects of order $a$. 
The reason is that beside our nonperturbative value for $\csw$ there are boundary improvement coefficients for which only
perturbative estimates are available. 
We use those to the known order \cite{Luscher:1993gh,Bode:1999sm,Sint:1995ch}
which ensures that $\mathcal{O}(a)$ cutoff effects in the step scaling function appear
only starting at three-loop order. 

The details of cutoff effects also depend on how the condition of a
massless scheme is exactly implemented at a finite lattice spacing. As in 
the $\nf=2$ computation, we define the massless point on the 
smaller of the pair of lattices entering the step scaling function. 
More precisely, the (unrenormalized) PCAC quark mass, 
\begin{equation}\label{mpcac}
 m(x_{0})=\frac{\frac{1}{2}(\drvstar0+\drv0)\fa(x_{0})+\ca a\drvstar0\drv0\fp(x_{0})}{2\fp(x_{0})}
\end{equation}
at $x_{0}=\frac{T}{2}$,
\begin{equation}\label{m1}
m_{1}=m\left(\frac{T}{2}\right)\,,
\end{equation}
is set to zero. The definition of the correlation functions $\fa,\fp$ is found e.g. in \cite{Tekin:2009,alpha:letter}. The second reference also contains a discussion of the influence of choosing the massless point in a way which differs from (\ref{m1}) at finite lattice spacing. For the improvement coefficient $\ca$ in (\ref{mpcac}) we used the 1-loop result \cite{Luscher:1996vw}.

In addition to the various improvement terms in the action
and $\ca$ in (\ref{mpcac}) we exploit our knowledge of $\delta_1$ and  $\delta_2$ from
the perturbative calculation of
\begin{equation}
 \delta (u,a/L)=\frac{\Sigma (u,a/L)-\sigma (u)}{\sigma (u)}=\delta_{1}(a/L)u+\delta_{2}(a/L)u^2+\dots,
\end{equation}
\begin{eqnarray}
 \delta_1(a/L)&=&\delta_{10}(a/L)+\delta_{11}(a/L)\nf\\
 \delta_2(a/L)&=&\delta_{20}(a/L)+\delta_{21}(a/L)\nf+\delta_{22}(a/L)\nf^2
\end{eqnarray}
with the coefficients taken from \cite{Gehrmann:2001yn}. For $\nf=4$ we have 
\begin{displaymath}
\begin{tabular}{ccl}\hline
$L/a$  & $\delta_{1}$ &$\quad\delta_{2}$ \\\hline\hline
$4$  &$-0.0102$ &$0.0073$\\
$6$  &$-0.0045$ &$0.0013$\\
$8$  &$-0.0024$ &$0.00013$\\\hline
\end{tabular} 
\end{displaymath}
Using these data we form the lattice step scaling function~\cite{alpha:SU2impr} 
\begin{equation}
 \Sigma^{(2)}(u,a/L)=\frac{\Sigma (u,a/L)}{1+\delta_{1}(a/L)u+\delta_{2}(a/L)u^2}
\end{equation}
which we expect to 
have smaller overall cutoff effects. They still start at order
$a\times u^4$ but terms of order $a^m\times u^n$ are removed for
all $m$ and for $n\leq3$ (in fact non-perturbatively in $a$). 
As mentioned previously, the order $a\times u^4$ terms are due to the 
only perturbatively known boundary improvement terms. Their influence
was explicitly checked for $\nf=2$ and found to be minor \cite{DellaMorte:2004bc}, 
such that also here we assume that  the step scaling function 
converges {\em effectively} at a rate
\begin{equation}
  \Sigma^{(2)}(u,a/L)=\sigma(u)+\mathcal{O}(a^2).
\end{equation}

Once the non-perturbatively determined continuum
step scaling function $\sigma(u)$ is known, the running of the coupling can be
computed easily. 
For this purpose, one solves the recurrence
\begin{equation}\label{recursion}
 u_{i}=\sigma(u_{i+1}),\quad i=0,\dots,n,\quad u_{0}=\umax=\gbar^2\left(\Lmax\right),
\end{equation}
where $\umax$ is chosen such that the corresponding scale $\Lmax$ is in the hadronic regime of QCD. Proceeding in this way the coupling can be obtained over a wide range of energies. At a sufficiently large energy $\mu=2^{k}/\Lmax$ ($k\lesssim n$), perturbation theory can be applied for determining the quantity $\Lambda\Lmax$ using (\ref{Lambda}) 
with the $\beta$ function truncated at 2-loop and 3-loop  respectively. 
\section{Numerical computation and results}
\subsection{Simulation parameters and raw data}
The choice of the improvement coefficients $\ct$ , $\ctildet$ and $\csw$ were
as discussed in our preceding paper \cite{Tekin:2009}. 
The range of $\beta$ for our simulations was limited by the validity range of
$\csw$ with four flavors ($\beta\le 5.0$) \cite{Tekin:2009}. Since our code
allows only an even number of lattice points in each direction and lattices
beyond $L/a=16$ are too time-consuming for our present resources, 
we chose lattices ($T=L$) $L/a=4,6,8$. We picked a sufficient number of values 
of $\beta \in [5.0 , 9.5]$ to adequately map out a range $\gbar^2(L)=0.9 \ldots 2.7$,
$\gbar^2(2L)=1.0 \ldots 3.5$. 
With some tuning of the hopping parameter $\kappa$ 
the quark mass was kept small enough ($|m_1 L| \leq 0.005$) such that 
mass-effects in the step scaling function are negligible.

We performed our simulations on 4-5 crates of the apeNEXT
machine at DESY Zeuthen
over a period of about a year. The raw data are listed in the appendix.
\subsection{Analysis of data}
\begin{figure}[htb]
$$\includegraphics[height=9cm]{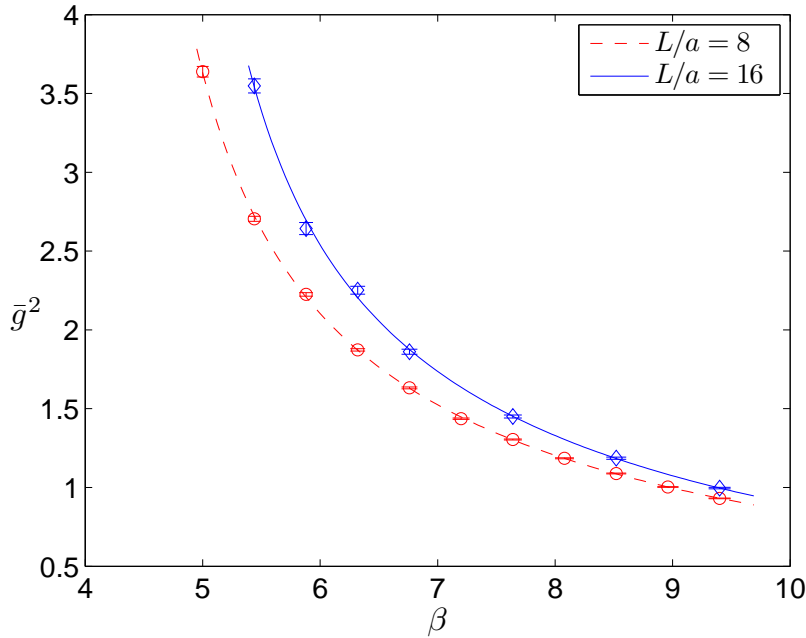}$$
\caption{Data points for $L/a=8$ and $L/a=16$ and their
  interpolations.
The parameter $n$ in (\ref{fitfunction}) is set to three. The hopping parameter $\kappa$ was tuned only on the small lattices $L/a=4,6,8$ such that the PCAC mass (\ref{m1}) vanished.}\label{gbarbeta}
\end{figure}
The computation of the step scaling function on the lattice as described in
section \ref{ssf_cont} requires $\gbar^2(L)$ to be fixed to certain values
$u$ while the resolution $a/L$ is changed. Previously this was realized 
by tuning $\beta$ for each pair $u,L/a$ \cite{Luscher:1993gh,mbar:pap1,DellaMorte:2004bc}.
Instead we here follow the more convenient proposal of \cite{Appelquist:2009ty} to 
pick a sufficient range and number of bare couplings for each considered $L/a$ 
and interpolate the running coupling $\gbar^2(\beta,L/a)$ with a smooth
function
of $\beta$. Afterwards
the function allows access to any value of $\gbar^2$ in the covered range.
As an interpolation we took
\begin{equation}\label{fitfunction}
\bar{g}^2(\beta,L/a)=\frac{6}{\beta}\left[\sum_{m=0}^{n}c_{m,L/a}\left(\frac{6}{\beta}\right)^m\right]^{-1}\,
\end{equation}
motivated by perturbation theory. Note, however, that we do not fix the 
known perturbative expansion coefficients; we do not even require
$c_{0,L/a}=1$. Somewhat different forms and values $n$ were checked and we
verified that our results do not depend on the details of these
interpolations. An example is shown in figure~\ref{gbarbeta} 
for a pair of lattice sizes, namely $L/a=8$ and
$L/a=16$. The coefficients  $c_{m,L/a}$
were determined by a standard $\chi^2$ fit.
\begin{table}[htb]
\centering
\begin{tabular}{C{1cm}C{0.5cm}p{1.7cm}p{2cm}C{1cm}C{0.5cm}p{2cm}p{1.9cm}}\hline
$u$ & $L/a$ & $\Sigma(u,a/L)$ & $\Sigma^{(2)}(u,a/L)$ & $u$ & $L/a$ & $\Sigma(u,a/L)$ & $\Sigma^{(2)}(u,a/L)$\\\hline
$0.93$   & $4$ & $0.995(2)$ & $0.999(2)$ & $1.4435$ & $4$ & $1.608\p(3)$ & $1.608\p(3)$\\ 
         & $6$ & $1.000(3)$ & $1.004(3)$ &          & $6$ & $1.627\p(6)$ & $1.633\p(6)$\\
         & $8$ & $0.995(5)$ & $0.997(5)$ &          & $8$ & $1.632\p(8)$ & $1.637\p(8)$\\
$1$      & $4$ & $1.076(2)$ & $1.079(2)$ & $1.6285$ & $4$ & $1.844\p(5)$ & $1.839\p(5)$\\
         & $6$ & $1.083(3)$ & $1.086(3)$ &          & $6$ & $1.868\p(8)$ & $1.875\p(8)$\\
         & $8$ & $1.079(4)$ & $1.081(4)$ &          & $8$ & $1.874(11)$ & $1.880(11)$\\
$1.0813$ & $4$ & $1.171(2)$ & $1.174(2)$ & $1.8700$ & $4$ & $2.169\p(8)$ & $2.155\p(8)$ \\
         & $6$ & $1.179(4)$ & $1.183(4)$ &          & $6$ & $2.199(13)$ & $2.208(13)$ \\
         & $8$ & $1.178(5)$ & $1.181(5)$ &          & $8$ & $2.203(17)$ & $2.212(17)$ \\
$1.1787$ & $4$ & $1.286(2)$ & $1.287(2)$ & $2.2003$ & $4$ & $2.650(11)$ & $2.617(11)$ \\
         & $6$ & $1.297(5)$ & $1.301(5)$ &          & $6$ & $2.688(17)$ & $2.698(17)$ \\
         & $8$ & $1.298(6)$ & $1.301(6)$ &          & $8$ & $2.684(24)$ & $2.697(24)$ \\
$1.2972$ & $4$ & $1.428(3)$ & $1.430(3)$ & $2.6870$ & $4$ & $3.462(22)$ & $3.378(21)$ \\
         & $6$ & $1.442(6)$ & $1.448(6)$ &          & $6$ & $3.507(40)$ & $3.517(40)$ \\
         & $8$ & $1.446(7)$ & $1.450(7)$ &          & $8$ & $3.477(44)$ & $3.496(44)$ \\\hline
\end{tabular}
\caption{Results for $\Sigma(u,a/L)$ and $\Sigma^{(2)}(u,a/L)$ for different lattices and couplings $u$.}\label{ssftable}
\end{table}
From the interpolated couplings we computed 
$\Sigma(u,a/L)$ for a number of values $u$ starting from an initial coupling
$u_{\textrm{initial}}=0.9$ and following {\em roughly} a sequence given by the recursion
(\ref{recursion}).

\subsubsection{Error propagation}
The uncertainties of the initial MC data which can be found in the appendix are 
statistically uncorrelated. For the purpose of propagating their errors,
let us collect them in a vector $x$ with components 
$x_i$. The above fit function ${\bar{g}^2(\beta,L/a)}$ may then be regarded
a function $f(x)$ of the initial data\footnote{We neglect that in the way
  we determine the interpolation of 
  $\bar{g}^2$ there is also a dependence on the uncertainties 
  $\delta x_i$. In the fit we
  could also replace the errors $\delta x_i$ by a smooth predefined function of $\beta$.
  The results would not change significantly.
}.
The error $\delta f$ of  $f$ is simply estimated by
\begin{equation}
 \left(\delta f\right)^2 = \sum_i \left(\frac{\partial f}{\partial x_i}\right)^2(\delta x_{i})^2\,.
\end{equation}
Since $f$ is a relatively complicated function, it is convenient to avoid
computing the derivatives ${\partial f}/{\partial x_i}$ analytically. 
Instead we just compute them by a symmetric finite difference,
with the step in $x_i$ given by $\delta x_{i}$ itself. 
This convenient method is applied for estimating the errors of
all quantities derived from our data in the following. If desired also the 
correlation matrix of the errors of different observables can be obtained this way.

\subsection{Results}
\begin{figure}[!p]
\vspace*{-12mm}
$$\includegraphics[width=14.3cm]{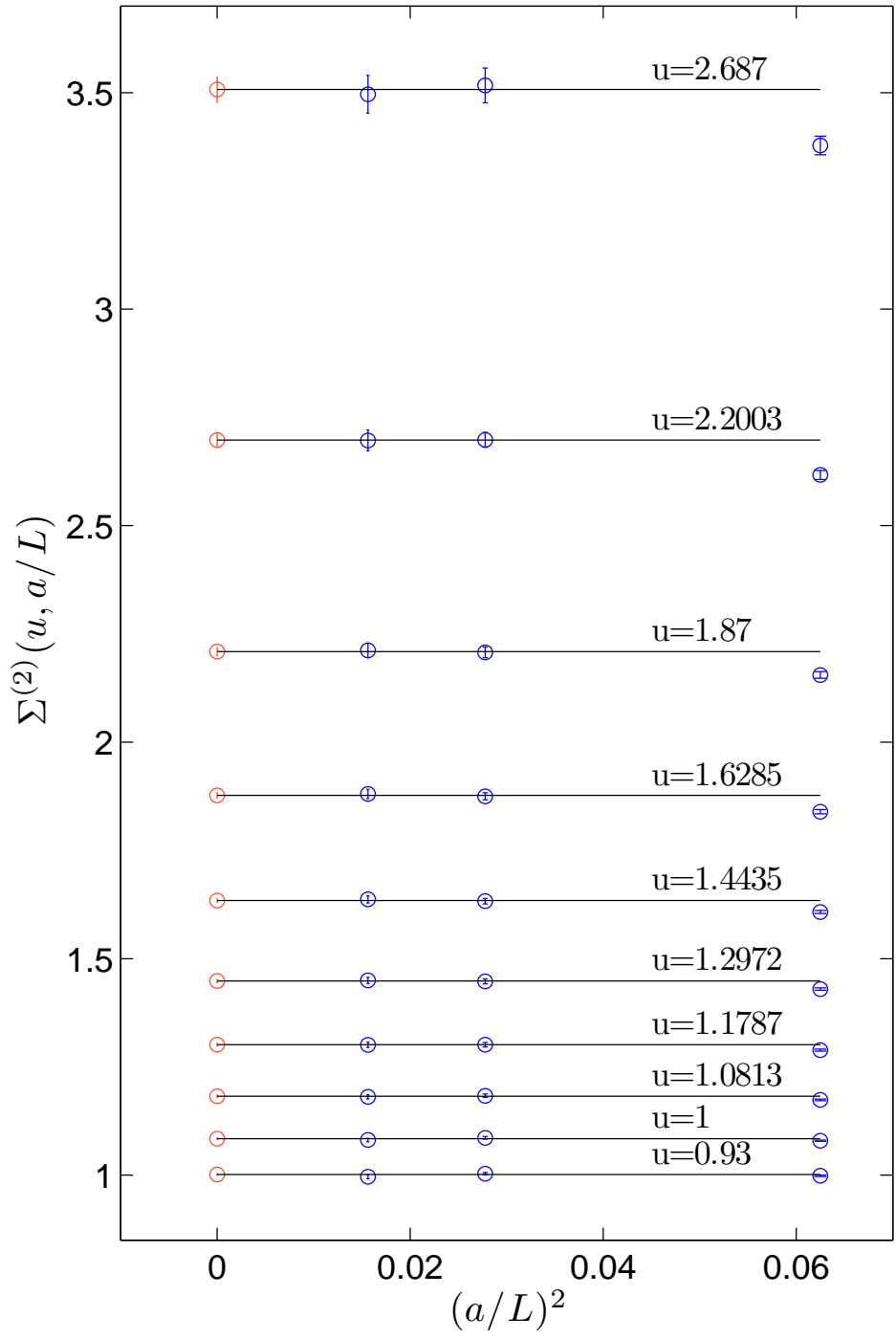}\vspace*{-12mm}$$
\captionof{figure}{Continuum extrapolation of the step scaling function (constant fit).}\label{continuum}
\end{figure}

Our result for both step scaling functions $\Sigma(u,a/L)$ and the perturbatively corrected $\Sigma^{(2)}(u,a/L)$ are listed in table~\ref{ssftable}.
As one can see in the visualization of our data, figure~\ref{continuum}, 
the cut-off effects seem to be very small except for $L/a=4$. 
As a precaution against higher order cutoff effects, we thus excluded the data set
of our coarsest lattice from our analysis leading to the continuum step
scaling function.  We carried out three different analysis.
\begin{itemize}
 \item {\em Constant fit:} A fit of $\Sigma^{(2)}(u,a/L)$ for $L/a=6,8$ to a constant, for
      each $u$. 
 \item {\em Global fit:} A fit
  \begin{equation} \label{imprsigma}
    \Sigma^{(2)}(u,a/L)=\sigma(u)+\rho\, u^4\,(a/L)^2.
  \end{equation}
  with a separate, independent parameter $\sigma(u)$ for
  each value $u$ but a common parameter $\rho$ modelling the cutoff-effects. 
 \item {\em $L/a=8$ data:} Using directly $\sigma(u)=\Sigma^{(2)}(u,1/8)$.
\end{itemize} 
The three different ans\"atze yield results which are in complete agreement with each other as
seen in Table~\ref{fittable}. The value of $\rho$ in (\ref{imprsigma}) comes out to be $\rho=0.007(85)$ which is a good indication that cutoff effects are negligible in the data for $L/a=6,8$.

In figure \ref{continuum} we depict the continuum extrapolation with the constant fit.
However, as our final results we take just the $L/a=8$ data. This is more
conservative and we are confident that the statistical errors dominate
over residual cutoff effects. In particular these
data agree with the $L/a=6$ data and also using $\Sigma(u,a/L)$ instead of $\Sigma^{(2)}(u,a/L)$ has a negligible effect.
\begin{table}[htb]
\begin{center}
\begin{tabular}{cccc}\hline
  $u$ &\multicolumn{3}{c}{$\sigma(u)$}\\
      & constant fit & global fit & $L/a=8$ data\\\hline
 $0.9300$ & $1.002\p(3)$ & $1.002\p(3)$ & $0.997\p(5)$\\        
 $1.0000$ & $1.084\p(3)$ & $1.084\p(3)$ & $1.081\p(4)$\\
 $1.0813$ & $1.182\p(3)$ & $1.182\p(4)$ & $1.181\p(5)$\\
 $1.1787$ & $1.301\p(4)$ & $1.301\p(5)$ & $1.301\p(6)$\\
 $1.2972$ & $1.448\p(5)$ & $1.448\p(7)$ & $1.450\p(7)$\\
 $1.4435$ & $1.634\p(5)$ & $1.634(10)$ & $1.637\p(8)$\\
 $1.6285$ & $1.877\p(7)$ & $1.877(16)$  & $1.880(11)$\\
 $1.8700$ & $2.209(10)$  & $2.207(27)$  & $2.212(17)$\\
 $2.2003$ & $2.698(14)$  & $2.694(49)$  & $2.697(24)$\\
 $2.6870$ & $3.507(30)$  & $3.50\p(10)$   & $3.496(44)$\\\hline
\end{tabular}
\caption{Results of different fit procedures (as described in the text) for the continuum extrapolation of the step scaling function.}\label{fittable}
\end{center}
\end{table}

\begin{figure}[htb]
 $$\includegraphics[height=10cm]{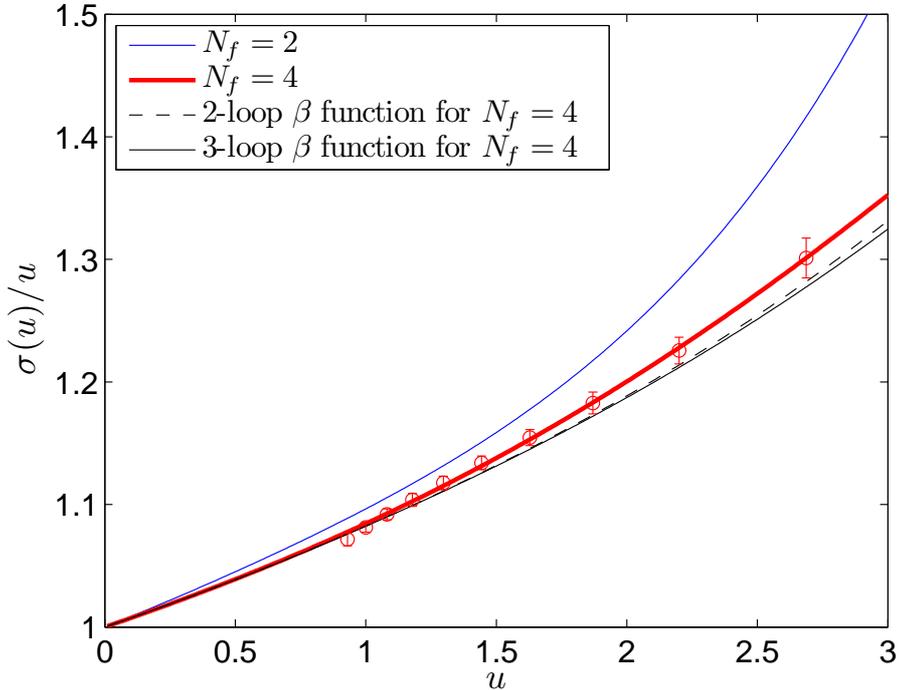}$$\vspace*{-12mm}
\caption{The step scaling function for $\Nf=2,4$ and the perturbative results. The thickest line is the fit of our data points. The upper solid line is the $\Nf=2$ result and the lower lines show the perturbative results.
}\label{ssf}
\end{figure}

Using a polynomial of degree five in $u$, we performed a constrained interpolation of the data in the fourth column in table \ref{fittable}. 
The coefficients up to $u^3$ were fixed by perturbation theory. Our fit  
\begin{equation}\label{ssfparam}
 \sigma (u)=u+s_{0}u^2+s_{1}u^3+0.0036\,u^4-0.0005\,u^5,\quad 0 \leq u \leq 2.7.
\end{equation}
is shown in figure \ref{ssf} (thick line). 
The perturbative step scaling functions are close to the one sigma range of the 
non-perturbative data points over the whole interval of the coupling $u$.
We note also a peculiarity. Perturbation theory at 3-loop lies below
the 2-loop truncation of the $\beta$-function and 
further away from the non-perturbative result. However, the 3-loop coefficient 
$b_2$ (eq. (\ref{b2})) in the \SF scheme changes its sign between $\Nf=2$ and $\Nf=3$
and it is rather small for $\Nf=4$. It is hence not unlikely that the 
4-loop term would move the perturbative curve closer again.

\begin{figure}[ht]
 $$\includegraphics[height=10cm]{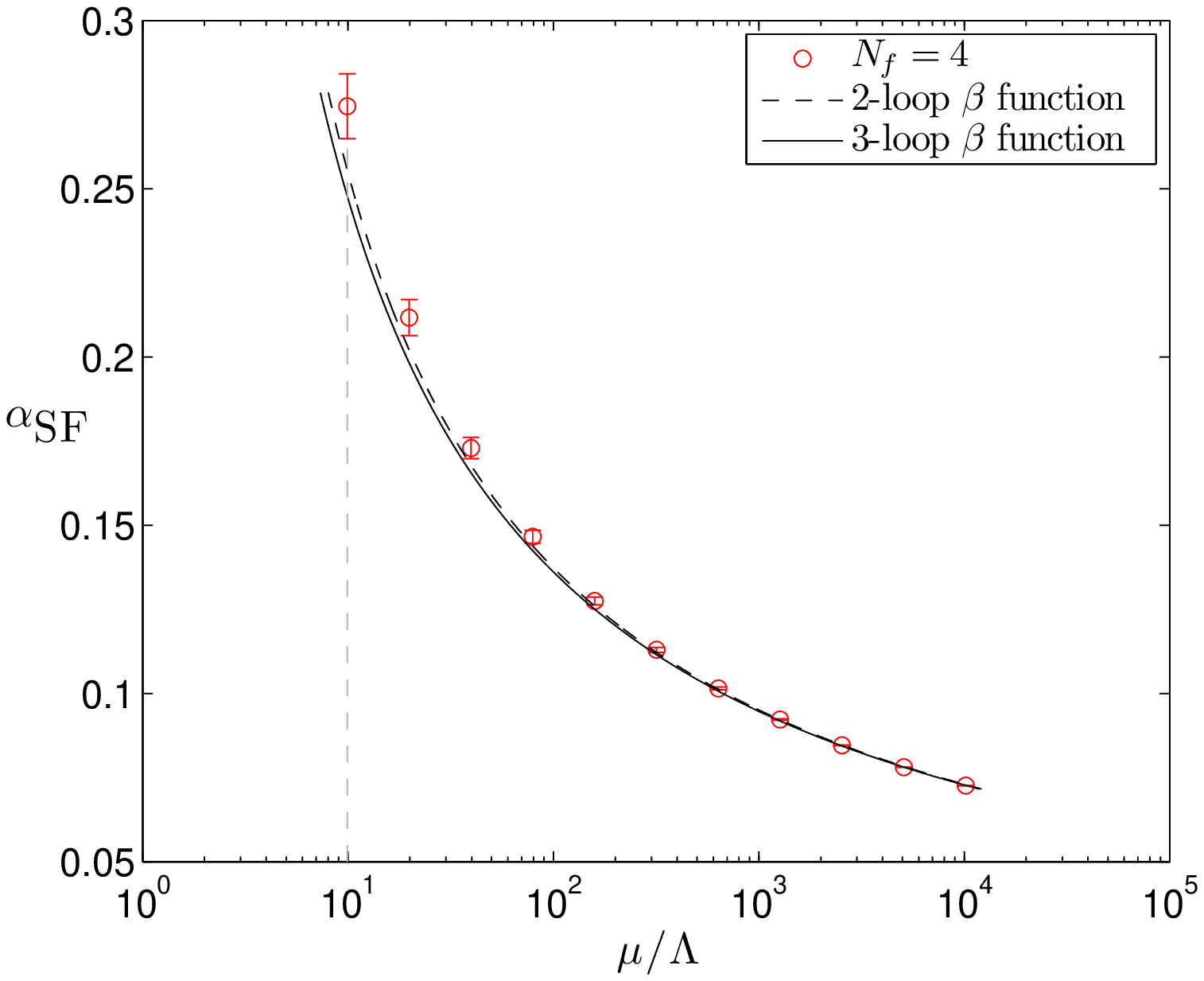}$$
 \caption{The running coupling in the \SF scheme. The gray vertical dashed line is only for the guidance of the eyes to show what the perturbation theory predicts for $\alphaSF$ at our lowest energy.}\label{running}
\end{figure}

Using the parameterization (\ref{ssfparam}) of the step scaling function, we calculated the
combination $\ln(\Lambda\Lmax)$ starting from the highest coupling $\umax=\gbar^2(\Lmax)$ which was covered by our non-perturbative step scaling function and solved the recursion step (\ref{recursion}) numerically to obtain the couplings $u_i$ which correspond to the energy scales $\mu=2^i/\Lmax$. With the help of (\ref{Lambda}) and using the truncated 3-loop $\beta$ function, we computed the values for $\ln(\Lambda\Lmax)$  recorded in table \ref{lambdatable}.
\begin{table}[htb]
\begin{center}
{\small
\begin{tabular}{ccccccc}\hline
& \multicolumn{2}{c}{constant fit} & \multicolumn{2}{c}{global fit} & \multicolumn{2}{c}{$L/a=8$ data}\\
$i$ & $u_i$ & $\ln(\Lambda\Lmax)$ & $u_i$ & $\ln(\Lambda\Lmax)$ & $u_i$ & $\ln(\Lambda\Lmax)$\\\hline
 $0$ & $3.45\p\p\p\p$           & $-2.028\p\p\p$         & $3.45\p\p\p\p$      & $-2.028\p\p\p$     & $3.45\p\p\p\p$     & $-2.028\p\p\p$\\     
 $1$ & $2.660(14)$      & $-2.074(17)$     & $2.666(46)$ & $-2.066(56)$ & $2.660\p(21)$  & $-2.073(26)$\\
 $2$ & $2.173(13)$      & $-2.117(24)$     & $2.179(45)$ & $-2.105(83)$ & $2.173\p(20)$  & $-2.116(37)$\\
 $3$ & $1.842(11)$      & $-2.155(28)$     & $1.847(37)$ & $-2.141(97)$ & $1.842\p(17)$  & $-2.153(44)$\\
 $4$ & $1.6013(90)$     & $-2.188(32)$     & $1.606(30)$ & $-2.17\p(10)$ & $1.602\p(14)$  & $-2.185(50)$\\
 $5$ & $1.4187(78)$     & $-2.217(35)$     & $1.422(25)$ & $-2.20\p(11)$ & $1.419\p(13)$  & $-2.213(56)$\\
 $6$ & $1.2748(70)$     & $-2.241(39)$     & $1.278(20)$ & $-2.23\p(11)$ & $1.275\p(11)$  & $-2.238(63)$\\
 $7$ & $1.1583(63)$     & $-2.263(43)$     & $1.161(17)$ & $-2.25\p(12)$  & $1.159\p(10)$  & $-2.259(70)$\\
 $8$ & $1.0620(58)$     & $-2.282(47)$     & $1.064(15)$ & $-2.27\p(12)$  & $1.0626(95)$ & $-2.278(76)$\\
 $9$ & $0.9809(53)$     & $-2.299(50)$     & $0.982(13)$ & $-2.29\p(12)$  & $0.9815(87)$ & $-2.294(83)$\\
 $10$ & $0.9117(49)$    & $-2.315(54)$     & $0.913(11)$ & $-2.30\p(12)$  & $0.9122(81)$ & $-2.309(89)$\\\hline
\end{tabular}
\caption{Results for $\ln(\Lambda\Lmax)$ with different fit strategies.}\label{lambdatable}
}
\end{center}
\end{table}
From the $L/a=8$ results we quote
\begin{equation}\label{lnlambdalmax}
 \ln(\Lambda\Lmax)=-2.294(83)\quad\text{ at }\umax=3.45
\end{equation}
as our final result. This determination of the $\Lambda$-parameter in units of $\Lmax$ has a precision of $\approx 8\%$. It remains to gauge $\Lmax$ in physical units
through a large volume computation. Therefore we here show the running of the coupling 
in the \SF scheme in units of $\Lambda$. 
Figure \ref{running} displays $\alphaSF$ computed from the sixth column of table \ref{lambdatable}. We observe that upon the iterative application of the step scaling
function the difference between the perturbative (using (\ref{lnlambdalmax})) 
and the non-perturbative coupling is 
around a 3-sigma effect at the strongest coupling.

\section{Conclusions}

We computed the step scaling function of the QCD coupling in the \SF 
scheme with four massless flavors. We used $\mathcal{O}(a)$ improved 
Wilson quarks after first determining the Sheikholeslami Wohlert coefficient
$\csw$ \cite{Tekin:2009}. The resulting cutoff effects are very small
(figure \ref{continuum}, table~\ref{ssftable}) allowing for a continuum extrapolation. While the data
are compatible with a constant for $L/a\geq6$, we assume this form only
for $L/a\geq8$; the smaller lattices thus only enter the analysis by
demonstrating that cutoff effects are small. We emphasize that this statement
refers to the present level of statistical errors. If in the future statistical
errors are further reduced, larger $L/a$ will be necessary at the same time.
It will be very interesting to see also the efficiency of computations
with different regularizations of the \SF as well as the corresponding
test of the universality of the continuum limit. Most notably there are chirally
rotated boundary conditions for the quarks \cite{SF:chirrot1,nara:stefan,lat09:jeni}
and staggered quarks \cite{SF:stagg,lat07:paula} for which results are
expected soon.

Already now, we do observe a small but significant deviation from 3-loop perturbation theory at the largest coupling reached in figure \ref{running}. It is about 10\% (three standard deviations) and the \SF coupling has a value of $\alphaSF \approx 0.28$. For $\nf=2$ a similar effect was visible only for larger coupling~\cite{DellaMorte:2004bc}\footnote{For $\alphaSF \approx 0.45$ a similar deviation is visible but there are no non-perturbative data points in between $\alphaSF \approx 0.28$ and $\alphaSF \approx 0.45$ to see better where this sets in.}. These findings underline the necessity of going to weak coupling before applying perturbation theory. 

Clearly the present work has brought us a good step closer to the computation
of the $\Lambda$-parameter in 4-flavor QCD, which may then be perturbatively connected to 
e.g. the 5-flavor $\MSbar$ coupling at the Z-pole. 
However, the technically introduced scale $\Lmax$ remains to be expressed in physical units
through large volume 4-flavor simulations. Apart from the challenge of 
tuning more parameters, one needs to treat a massive charm quark at small 
enough lattice spacing. Presently this appears to be a considerable challenge due to
a severe slowing down of lattice simulation algorithms at small lattice spacings
\cite{DelDebbio:2002xa,lat09:stefan}.

\section*{Acknowledgements}
This work is part of the ALPHA-collaboration research programme. We thank NIC
for allocating computer time on the APE computers at DESY, Zeuthen and the
staff of the computer center at Zeuthen for their support. This work is
supported by the Deutsche Forschungsgemeinschaft (DFG) in the framework of
SFB/TR 09 and by the European Community
through EU Contract No.~MRTN-CT-2006-035482, ``FLAVIAnet''.
\appendix
\section*{Appendix: The MC data}
We here list the results of our MC simulations. Each row refers to a separate simulation of about $50000$ trajectories unless otherwise noted. Some of these simulations consist of independent replica (between 1 and 16). Measurements were taken after every trajectory for which we chose trajectory
length $\tau= 1$.
\begin{center}
\begin{tabular}{|c|c||r|r|r|r|}\hline
        & &\multicolumn{2}{|c|}{$L/a=4$}&\multicolumn{2}{|c|}{$L/a=8$}\\\hline\hline
$\beta$ & $\kappa$   & $\bar{g}^2$   & $am_1$         & $\bar{g}^2$   & $am_1$\\\hline
$5.0$   & $0.137975$ & $2.913\p(10)$ & $0.00033(42)$  & $3.932\p(39)$  & $0.03752\p(13)$\\\hline
$5.3$   & $0.137110$ & $2.4700\p(76)$  & $0.00040(34)$  & $3.049\p(22)$  & $0.02934\p(10)$\\\hline
$5.6$   & $0.136371$ & $2.1505\p(49)$  & $0.00042(29)$  & $2.575\p(15)$  & $0.024981(80)$\\\hline
$6.2$   & $0.135082$ & $1.7300\p(31)$  & $0.00162(24)$  & $1.9853(91)$ & $0.020192(61)$\\\hline
$6.8$   & $0.134053$ & $1.4556\p(21)$  & $0.00080(20)$  & $1.6256(61)$ & $0.016432(52)$\\\hline
$7.4$   & $0.133188$ & $1.2609\p(13)$  & $0.00032(18)$  & $1.3844(42)$ & $0.013977(47)$\\\hline
$8.0$   & $0.132455$ & $1.1119\p(10)$  & $0.00070(15)$  & $1.2074(32)$ & $0.012435(40)$\\\hline
$8.6$   & $0.131860$ & $0.99575(77)$ & $-0.00021(14)$ & $1.0678(25)$ & $0.010382(35)$\\\hline
$9.2$   & $0.131309$ & $0.90315(54)$ & $0.00094(11)$  & $0.9662(20)$ & $0.010176(33)$\\\hline
\end{tabular}
\captionof{table}{The raw data for $L/a=4$ and $L/a=8$.}
\end{center}
\begin{center}
\begin{tabular}{|c|c||r|r|r|r|}\hline
        & &\multicolumn{2}{|c|}{$L/a=6$}&\multicolumn{2}{|c|}{$L/a=12$}\\\hline\hline
$\beta$ & $\kappa$   & $\bar{g}^2$   & $am_1$         & $\bar{g}^2$   & $am_1$\\\hline
 $5.25$  & $0.138027$ & $2.749\p(13)$  & $-0.00005\p(16)$  & $3.635\p(46)$  & $0.000929(56)$\\\hline
 $5.55$  & $0.137173$ & $2.3507(92)$ & $0.00110\p(13)$   & $2.904\p(29)$  & $0.000704(43)$\\\hline
 $5.85$  & $0.136443$ & $2.0865(71)$ & $0.00053\p(11)$   & $2.529\p(23)$  & $-0.000031(37)$\\\hline
 $6.45$  & $0.135190$ & $1.6948(46)$ & $-0.000294(94)$ & $1.953\p(14)$  & $-0.000922(31)$\\\hline
 $7.05$  & $0.134123$ & $1.4361(32)$ & $0.000488(78)$  & $1.6211(88)$ & $-0.000227(25)$\\\hline
 $7.65$  & $0.133261$ & $1.2500(24)$ & $0.000437(69)$  & - & -\\\hline
 $8.25$  & $0.132538$ & $1.1025(18)$ & $0.000435(62)$  & $1.2051(50)$ & $-0.000347(20)$\\\hline
 $8.85$  & $0.131935$ & $0.9908(14)$ & $0.000154(57)$  & - & -\\\hline
 $9.45$  & $0.131411$ & $0.8975(12)$ & $0.000237(51)$  & $0.9628(31)$ & $-0.000547(16)$\\\hline
\end{tabular}
\captionof{table}{The raw data for $L/a=6$ and $L/a=12$.}
\end{center}
\newpage
\begin{center}
\begin{tabular}{|c|c||r|r|r|r|}\hline
        & &\multicolumn{2}{|c|}{$L/a=8$}&\multicolumn{2}{|c|}{$L/a=16$}\\\hline\hline
$\beta$ & $\kappa$   & $\bar{g}^2$   & $am_1$         & $\bar{g}^2$   & $am_1$\\\hline
 $5.0$   & $0.138910$ & $3.638\p(34)$  & $0.00037\p(14)$  & - & -\\\hline
 $5.44$  & $0.137507$ & $2.705\p(16)$  & $0.000640(83)$ & $3.548\p(45)$  & $-0.000872(23)$\\\hline
 $5.88$  & $0.136393$ & $2.225\p(11)$  & $0.000306(66)$ & $2.643\p(38)$  & $-0.001120(25)$\\\hline
 $6.32$  & $0.135433$ & $1.8728(77)$ & $0.000288(57)$ & $2.252\p(25)$  & $-0.000875(22)$\\\hline
 $6.76$  & $0.134597$ & $1.6319(56)$ & $0.000748(58)$ & $1.861\p(16)$  & $-0.000350(18)$\\\hline
 $7.2$   & $0.133903$ & $1.4364(42)$ & $0.000041(44)$ & - & -\\\hline
 $7.64$  & $0.133275$ & $1.3046(35)$ & $0.000233(40)$ & $1.4502(94)$ & $-0.000666(15)$\\\hline
 $8.08$  & $0.132736$ & $1.1852(29)$ & $0.000069(38)$ & - & -\\\hline
 $8.52$  & $0.132249$ & $1.0886(24)$ & $0.000328(36)$ & $1.1860(67)$ & $-0.000552(12)$\\\hline
 $8.96$  & $0.131821$ & $1.0034(20)$ & $0.000368(33)$ & - & -\\\hline
 $9.4$   & $0.131442$ & $0.9308(17)$ & $0.000284(32)$ & $0.9961(48)$ & $-0.000504(11)$\\\hline
\end{tabular}
\captionof{table}{The raw data for $L/a=8$ and $L/a=16$. The run $L/a=16$, $\beta=5.44$ has 98000 trajectories.}
\end{center}
\newpage


\begin{thebibliography}{10}

\bibitem{Butler:1994}
F.~Butler, H.~Chen, J.~Sexton, A.~Vaccarino, and D.~Weingarten.
\newblock {Hadron masses from the valence approximation to lattice QCD}.
\newblock {\em Nucl. Phys.}, B430:179, 1994, hep-lat/9405003.

\bibitem{spectrum:cppacs_nf0}
S.~Aoki et~al.
\newblock {Quenched light hadron spectrum}.
\newblock {\em Phys.Rev.Lett.}, 84:238, 2000, hep-lat/9904012.

\bibitem{spectrum:cppacs_nf2}
A.~Ali~Khan et~al.
\newblock {Light hadron spectroscopy with two flavors of dynamical quarks on
  the lattice}.
\newblock {\em Phys.Rev.}, D65:054505, 2002, hep-lat/0105015.

\bibitem{Aoki:2008sm}
S.~Aoki et~al.
\newblock {2+1 Flavor Lattice QCD toward the Physical Point}.
\newblock {\em Phys. Rev.}, D79:034503, 2009, arXiv:0807.1661.

\bibitem{spectrum:etmc_nf2_baryons}
C.~Alexandrou et~al.
\newblock {Light baryon masses with dynamical twisted mass fermions}.
\newblock {\em Phys.Rev.}, D78:014509, 2008, arXiv:0803.3190.

\bibitem{Durr:2008zz}
{S. D\"urr et al}.
\newblock {Ab-Initio Determination of Light Hadron Masses}.
\newblock {\em Science}, 322:1224, 2008, arXiv:0906.3599.

\bibitem{Luscher:1998pe}
{Martin L\"uscher}.
\newblock {Advanced lattice QCD}.
\newblock {\em Les Houches 1997, Probing the standard model of particle
  interactions, Pt. 2}, 1998, hep-lat/9802029.

\bibitem{Luscher:1993gh}
{Martin L\"uscher, Rainer Sommer, Peter Weisz, and Ulli Wolff}.
\newblock {A Precise determination of the running coupling in the SU(3)
  Yang-Mills theory}.
\newblock {\em Nucl. Phys.}, B413:481, 1994, hep-lat/9309005.

\bibitem{mbar:pap1}
Stefano Capitani, Martin {L\"uscher}, Rainer Sommer, and Hartmut Wittig.
\newblock Non-perturbative quark mass renormalization in quenched lattice
  {QCD}.
\newblock {\em Nucl. Phys.}, B544:669, 1999, hep-lat/9810063.

\bibitem{Takeda:2004}
S.~Takeda et~al.
\newblock {A scaling study of the step scaling function in SU(3) gauge theory
  with improved gauge actions}.
\newblock {\em Phys. Rev.}, D70:074510, 2004, hep-lat/0408010.

\bibitem{DellaMorte:2004bc}
Michele Della~Morte et~al.
\newblock {Computation of the strong coupling in QCD with two dynamical
  flavours}.
\newblock {\em Nucl. Phys.}, B713:378, 2005, hep-lat/0411025.

\bibitem{alpha:cppacs_nf3}
S.~Aoki et~al.
\newblock {Precise determination of the strong coupling constant in N(f) = 2+1
  lattice QCD with the Schr\"odinger functional scheme}.
\newblock {\em JHEP}, 0910:053, 2009, arXiv:0906.3906.

\bibitem{Appelquist:2009ty}
Thomas Appelquist, George~T. Fleming, and Ethan~T. Neil.
\newblock {Lattice Study of Conformal Behavior in SU(3) Yang-Mills Theories}.
\newblock {\em Phys. Rev.}, D79:076010, 2009, arXiv:0901.3766.

\bibitem{Mason:2005zx}
Q.~Mason et~al.
\newblock {Accurate determinations of $\alpha_s$ from realistic lattice QCD}.
\newblock {\em Phys. Rev. Lett.}, 95:052002, 2005, hep-lat/0503005.

\bibitem{Gockeler:2005rv}
{M. G\"ockeler et al}.
\newblock {A determination of the Lambda parameter from full lattice QCD}.
\newblock {\em Phys. Rev.}, D73:014513, 2006, hep-ph/0502212.

\bibitem{Shintani:2008ga}
E.~Shintani et~al.
\newblock {Lattice study of the vacuum polarization function and determination
  of the strong coupling constant}.
\newblock {\em Phys. Rev.}, D79:074510, 2009, arXiv:0807.0556.

\bibitem{Tekin:2009}
Fatih Tekin, Rainer Sommer, and Ulli Wolff.
\newblock {Symanzik improvement of lattice QCD with four flavors of Wilson
  quarks}.
\newblock {\em Phys. Lett.}, B683:75, 2010, arXiv:0911.4043.

\bibitem{Luscher:1991wu}
{Martin L\"uscher, Peter Weisz, and Ulli Wolff}.
\newblock {A Numerical method to compute the running coupling in asymptotically
  free theories}.
\newblock {\em Nucl. Phys.}, B359:221, 1991.

\bibitem{ParticleDataGroup}
Particle~Data Group.
\newblock {\em Particle Physics Booklet}.
\newblock http://pdg.lbl.gov, 2008.

\bibitem{Luscher:1992an}
{Martin L\"uscher, Rajamani Narayanan, Peter Weisz, and Ulli Wolff}.
\newblock {The Schr\"odinger functional: A Renormalizable probe for non-Abelian
  gauge theories}.
\newblock {\em Nucl. Phys.}, B384:168, 1992, hep-lat/9207009.

\bibitem{Sint:1993un}
Stefan Sint.
\newblock {On the Schr\"odinger functional in QCD}.
\newblock {\em Nucl. Phys.}, B421:135, 1994, hep-lat/9312079.

\bibitem{Sint:1995rb}
Stefan Sint.
\newblock {One loop renormalization of the QCD Schr\"odinger functional}.
\newblock {\em Nucl. Phys.}, B451:416, 1995, hep-lat/9504005.

\bibitem{Sint:1995ch}
Stefan Sint and Rainer Sommer.
\newblock {The Running coupling from the QCD Schr\"odinger functional: A One
  loop analysis}.
\newblock {\em Nucl. Phys.}, B465:71, 1996, hep-lat/9508012.

\bibitem{Bode:1999sm}
Achim Bode, Peter Weisz, and Ulli Wolff.
\newblock {Two loop computation of the Schr\"odinger functional in lattice
  QCD}.
\newblock {\em Nucl. Phys.}, B576:517, 2000, hep-lat/9911018.
\newblock Erratum-ibid.B600:453,2001, Erratum-ibid.B608:481,2001.

\bibitem{Sommer:2006sj}
Rainer Sommer.
\newblock {Non-perturbative QCD: Renormalization, $\mathcal{O}(a)$-improvement
  and matching to heavy quark effective theory}.
\newblock {\em hep-lat/0611020}, 2006, hep-lat/0611020.

\bibitem{Gehrmann:2001yn}
Bernd Gehrmann, Juri Rolf, Stefan Kurth, and Ulli Wolff.
\newblock {Schr\"odinger functional at negative flavour number}.
\newblock {\em Nucl. Phys.}, B612:3, 2001, hep-lat/0106025.

\bibitem{alpha:letter}
Achim Bode et~al.
\newblock First results on the running coupling in {QCD} with two massless
  flavors.
\newblock {\em Phys. Lett.}, B515:49, 2001, hep-lat/0105003.

\bibitem{Luscher:1996vw}
{M. L\"uscher and P. Weisz}.
\newblock {$\mathcal{O}(a)$ improvement of the axial current in lattice QCD to
  one-loop order of perturbation theory}.
\newblock {\em Nucl. Phys.}, B479:429, 1996, hep-lat/9606016.

\bibitem{alpha:SU2impr}
Giulia de~Divitiis et~al.
\newblock Universality and the approach to the continuum limit in lattice gauge
  theory.
\newblock {\em Nucl. Phys.}, B437:447, 1995, hep-lat/9411017.

\bibitem{SF:chirrot1}
Stefan Sint.
\newblock {The Schr\"odinger functional with chirally rotated boundary
  conditions}.
\newblock {\em PoS}, LAT2005:235, 2006, hep-lat/0511034.

\bibitem{nara:stefan}
Stefan Sint.
\newblock {Lattice QCD with a chiral twist}.
\newblock {\em hep-lat/0702008}, 2007, hep-lat/0702008.

\bibitem{lat09:jeni}
Jenifer~Gonzalez Lopez, Karl Jansen, Dru~B. Renner, and Andrea Shindler.
\newblock {Chirally rotated Schr\"odinger functional: non-perturbative tuning
  in the quenched approximation}.
\newblock {\em PoS}, LAT2009:199, 2009, arXiv:0910.3760.

\bibitem{SF:stagg}
Urs~M. Heller.
\newblock {The Schr\"odinger functional running coupling with staggered
  fermions}.
\newblock {\em Nucl.Phys.}, B504:435, 1997, hep-lat/9705012.

\bibitem{lat07:paula}
Paula Perez-Rubio and Stefan Sint.
\newblock {The SF running coupling with four flavours of staggered quarks}.
\newblock {\em PoS}, LAT2007:249, 2007, arXiv:0710.0583.

\bibitem{DelDebbio:2002xa}
Luigi Del~Debbio, Haralambos Panagopoulos, and Ettore Vicari.
\newblock {Theta dependence of SU(N) gauge theories}.
\newblock {\em JHEP}, 0208:044, 2002, hep-th/0204125.

\bibitem{lat09:stefan}
Stefan Schaefer, Rainer Sommer, and Francesco Virotta.
\newblock {Investigating the critical slowing down of QCD simulations}.
\newblock {\em PoS}, LAT2009, 2009, arXiv:0910.1465.

\end{thebibliography}
\end{document}